\documentstyle[epsfig,aps,pra]{revtex}
\begin{document}
\draft
\title{Direct and indirect strategies for phase measurement}
\author{M\'arcia T. Fontenelle\thanks{Present address: Laser Physics and Quantum Optics,
        Royal Institute of Technology, Lindstedtsv\"agen 24, 10044 Stockholm, Sweden.},
        Samuel L. Braunstein\thanks{Permanent address: SEECS, University of Wales,
                                Bangor, Gwynedd LL57 1UT, UK.},
        Wolfgang P. Schleich\thanks{Also at Max-Planck Institut f\"ur Quantenoptik,
                              85740 Garching, Germany}, and}
\address{Abteilung f\"ur Quantenphysik, Universit\"at Ulm, 89069 Ulm,
         Germany}
\author{Mark Hillery}
\address{Dept. of Physics, Hunter College of CUNY, 695 Park Ave. New York,
         NY 10021, USA}
\date{\today}
\maketitle
\begin{abstract}
Recently, Torgerson and Mandel [Phys.  Rev.  Lett.  \textbf{76}, 3939 (1996)]
have reported a disagreement between two schemes for measuring the phase
difference of a pair of optical fields.  We analyze these schemes and derive
their associated phase-difference probability distributions, including both their
strong and weak field limits.  Our calculation confirms the main point of
Torgerson and Mandel of the non-uniqueness of an operational definition of the
phase distribution. We further discuss the role of postselection of data and
argue that it cannot meaningfully improve the sensitivity.
\end{abstract}
\pacs{42.50.Dv,03.65.Bz}
%
%
%
%\narrowtext
%
%%%%%%%%%%%%%%%%%%%%%%%%%%%%%%%%%%%%%%%%%%%%%%%%%%%%%%%%%%%%%%%%%%%%%%%%%%%%
\section{Introduction}
\label{sec:introd}
%%%%%%%%%%%%%%%%%%%%%%%%%%%%%%%%%%%%%%%%%%%%%%%%%%%%%%%%%%%%%%%%%%%%%%%%%%%%

Lack of a canonical pair for the number and phase operators, $\hat n$ and $\hat
\varphi$, has led to much debate in quantum mechanics for many years
now~\cite{review}.  A more recent approach to the phase question involves
concentrating more on what the experimentalist actually does:  If his goal is a
precision measurement then he can perform ``any'' measurement followed by data
analysis to extract a classical parameter~\cite{Shapiro}, for instance phase
shift in an interferometer.  Alternatively, he may mentally lump together the
measurement and data analysis to construct an \emph{operational phase operator}
for his specific setup.  This latter approach has been championed by Mandel and
his coworkers \cite{Noh91,Noh92,Noh92a,Noh93a}. The explicit construction and
investigation of the properties of such phase observables can give us insight
into the nature of quantum states.

Recently, Torgerson and Mandel~\cite{Torgerson96} have compared two
schemes for measuring the phase-difference between a pair of
optical fields.  They found that a direct scheme, where a signal is
beat against a second one, and an indirect scheme, where the two
signals are beat against a common local oscillator, yield {\it
different\/} probability distributions for the measured
phase-difference.  In particular, they found that the schemes gave
radically different distributions for very weak signals.  Torgerson
and Mandel take the conflicting results as evidence of the
non-uniqueness of quantum phase.  In their analyses ambiguous data
is discarded.  This postselection procedure has generated some
discussion in the literature~\cite{Hradil93,Barnett}.  We analyze
these two schemes in the absence and presence of this postselection
and discuss its interpretation.  We study limits for both strong
and weak signals in these schemes and give closed form expressions
for the phase-difference probability distributions.

%%%%%%%%%%%%%%%%%%%%%%%%%%%%%%%%%%%%%%%%%%%%%%%%%%%%%%%%%%%%%%%%%%%%%%%%%%%%
\section{The NFM method}
\label{sec:NFM}
%%%%%%%%%%%%%%%%%%%%%%%%%%%%%%%%%%%%%%%%%%%%%%%%%%%%%%%%%%%%%%%%%%%%%%%%%%%%

In the present section we review the operational definition of
quantum phase based on the eight-port homodyne detector as well as its
photon count statistics.

%%%%%%%%%%%%%%%%%%%%%%%%%%%%%%%%%%%%%%%%%%%%%%%%%%%%%%%%%%%%%%%%%%%%%%%%%%%%
\subsection{NFM phase operators}
\label{sec:operators}
%%%%%%%%%%%%%%%%%%%%%%%%%%%%%%%%%%%%%%%%%%%%%%%%%%%%%%%%%%%%%%%%%%%%%%%%%%%%

\begin{figure}
\centerline{\epsfig{file=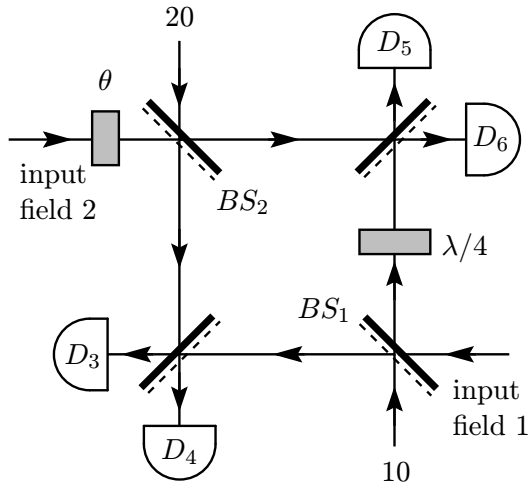}}\vspace{1ex}
\caption{Eight-port homodyne detector with the four input ports 1, 10,
2, and 20 and four output ports with detectors $D_3$, $D_4$, $D_5$,
and $D_6$ measuring the photon count differences $n_{43}=n_4-n_3$ and
$n_{65}=n_6-n_5$.  A quarter-wave plate causes a phase shift of $\pi/2$ in one
arm of the interferometer.  Input ports 10 and 20 remain open
corresponding to a vacuum input.  The input field 2 is phase
shifted by an angle $\theta$. A measurement of photon count
differences is made for each value of $\theta$.}
\label{fig:direct}
\end{figure}

The idea of Noh, Foug\`eres and Mandel of operationally defined
phase operators is guided by a classical analysis~\cite{Mandel92}
of the eight-port homodyne interferometer~\cite{Walker} shown in
Fig.~\ref{fig:direct}.  Replacing the classical light intensities
at the four detectors $D_i, (i=3,4,5,6)$ by number operators
$\hat{n}_i=\hat{a}^\dagger_i \hat{a}_i$, NFM propose
\cite{Noh91,Noh92,Noh92a,Noh93a,Mandel92} the phase operators
\begin{eqnarray}
\hat{C}_M&\equiv&\frac{\hat{n}_4-\hat{n}_3}
  {\sqrt{(\hat{n}_4-\hat{n}_3)^2 + (\hat{n}_6-\hat{n}_5)^2}}
  \equiv\frac{\hat{n}_{43}}{\sqrt{\hat{n}_{43}^2 + \hat{n}_{65}^2}}
  \label{eq:NFM-cos} \\
\noalign{\hbox{\rm and}}
\hat{S}_M&\equiv&\frac{\hat{n}_6-\hat{n}_5}
  {\sqrt{(\hat{n}_4-\hat{n}_3)^2 + (\hat{n}_6-\hat{n}_5)^2}}
  \equiv \frac{\hat{n}_{65}}{\sqrt{\hat{n}_{43}^2 +\hat{n}_{65}^2}}.
  \label{eq:NFM-sin}
\end{eqnarray}
In the classical limit the operators $\hat{C}_M$ and $\hat{S}_M$ become
\cite{Mandel92} the cosine and sine of the phase difference between two
classical electromagnetic fields in the modes 1 and 2.  Hence we can
consider these operators to be the extension of this classical
description of phase into the quantum domain.

How can we calculate expectation values of a function $f$ of these
operators $\hat{C}_M$ and $\hat{S}_M$?  According to the
NFM-prescription this expectation value reads
\begin{equation}
  \langle f(\hat{C}_M,\hat{S}_M) \rangle \equiv
  {\cal N} \sum_{n_{43},n_{65} \neq 0} f
  \left(\frac{n_{43}}{\sqrt{n_{43}^2+n_{65}^2}},
  \frac{n_{65}}{\sqrt{n_{43}^2+n_{65}^2}}\right)  W(n_{43},n_{65}),
  \label{eq:f-exp}
\end{equation}
where $W(n_{43},n_{65})$ denotes the joint count probability for
the differences $n_{43}\equiv n_4-n_3$ and $n_{65}\equiv n_6-n_5$
at the four detectors.  The sum in above equation is performed over
all $n_{43}$ and $n_{65}$, except $n_{43}=n_{65}=0$. In order to
avoid the problems associated with the definitions
(\ref{eq:NFM-cos}) and (\ref{eq:NFM-sin}) for $n_{43}=n_{65}=0$,
NFM disregard such measurements and renormalize the joint count
probability via ${\cal N}=[1-W(0,0)]^{-1}$.  The theoretical
results obtained by NFM using this formalism are in very nice
agreement~\cite{Noh91,Noh92,Noh92a,Noh93a} with their experiments.

Equation (\ref{eq:f-exp}) shows that the central quantity in the NFM
approach is the joint count
probability~\cite{Freyberger93,Freyberger93a,Freyberger93b}
\begin{equation}
W(n_{43},n_{65})=\mathop{{\sum}'}_{n_3,n_4,n_5,n_6}
\langle n_3,n_4,n_5,n_6|\hat\rho_{\rm out}|n_3,n_4,n_5,n_6\rangle,
\label{eq:joint-def}
\end{equation}
where $\hat\rho_{\rm out}$ is the density operator of the four input modes of the
interferometer after transformation through the beam splitters and the
$\lambda/4$ plate.  The ket $|n_3,n_4,n_5,n_6\rangle$ denotes the product number
state corresponding to the four output modes and the summation $\sum'$ is
performed for fixed differences $n_{43}$ and $n_{65}$.

%%%%%%%%%%%%%%%%%%%%%%%%%%%%%%%%%%%%%%%%%%%%%%%%%%%%%%%%%%%%%%%%%%%%%%%%%%%%
\subsection{Eight-port homodyne count statistics}
\label{sec:8-port}
%%%%%%%%%%%%%%%%%%%%%%%%%%%%%%%%%%%%%%%%%%%%%%%%%%%%%%%%%%%%%%%%%%%%%%%%%%%%

We now consider the case in which the input density matrix is expressed
in terms of coherent states, that is
\begin{equation}
\hat\rho_{\rm in}=\int d^2\beta_1 d^2\beta_2 P(\beta_1,\beta_2)
          |\beta_1\rangle_{1\;1}\langle\beta_1|
           \otimes|0\rangle_{10\;10}\langle0|
           \otimes|\beta_2\rangle_{2\;2}\langle\beta_2|
           \otimes|0\rangle_{20\;20}\langle0|,
\label{eq:rho-in}
\end{equation}
where $P(\beta_1,\beta_2)$ is a two-mode Glauber-Sudarshan distribution for the
fields entering ports 1 and 2. The output density operator just before
photodetection reads
\begin{eqnarray}
\hat\rho_{\rm out}&=&\int d^2\beta_1 d^2\beta_2 P(\beta_1,\beta_2)
\left|{\textstyle{\frac{1}{2}}}(\beta_1-\beta_2 e^{i\theta})\right\rangle_{3\;3}
       \left\langle{\textstyle{\frac{1}{2}}}(\beta_1-\beta_2 e^{i\theta})\right|
\nonumber\\
&&\qquad{}\otimes
\left|{\textstyle{\frac{1}{2}}}(\beta_1+\beta_2 e^{i\theta})\right\rangle_{4\;4}
        \left\langle{\textstyle{\frac{1}{2}}}(\beta_1+\beta_2 e^{i\theta})\right|
\nonumber\\
&&{}\qquad\otimes
\left|{\textstyle{\frac{1}{2}}}(-i\beta_1+\beta_2 e^{i\theta})\right\rangle_{5\:5}
        \left\langle{\textstyle{\frac{1}{2}}}(-i\beta_1+\beta_2 e^{i\theta})\right|
\nonumber\\
&&{}\qquad\otimes
\left|{\textstyle{\frac{1}{2}}}(-i\beta_1-\beta_2 e^{i\theta})\right\rangle_{6\;6}
        \left\langle{\textstyle{\frac{1}{2}}}(-i\beta_1-\beta_2 e^{i\theta})\right|,
\label{eq:rho-out}
\end{eqnarray}
where the phase shift $\theta$ was introduced into the field entering port 2 for
reasons which will become clear in the next section.

We calculate the joint count probability for the differences $n_{43}$ and
$n_{65}$ in the photocount distribution, for a given phase shift $\theta$,
by substituting Eq.~(\ref{eq:rho-out}) into Eq.~(\ref{eq:joint-def}), and
obtain
\begin{equation}
W(n_{43},n_{65}| e^{i\theta})=\int d^2\beta_1 d^2\beta_2 P(\beta_1,\beta_2)
K_{43}K_{65},
\label{eq:joint}
\end{equation}
where the kernels $K_{43}$ and $K_{65}$ are given by
\begin{eqnarray}
K_{43}&=&\mathop{{\sum}'}_{n_3,n_4}
\left|\langle n_3|{\textstyle{\frac{1}{2}}}(\beta_1-\beta_2e^{i\theta})\rangle\right|^2
\left|\langle n_4|{\textstyle{\frac{1}{2}}}(\beta_1+\beta_2e^{i\theta})\rangle\right|^2\\
\noalign{\hbox{\rm and}}
K_{65}&=&\mathop{{\sum}'}_{n_5,n_6}
\left|\langle n_5|{\textstyle{\frac{1}{2}}}(-i\beta_1+\beta_2e^{i\theta})\rangle\right|^2
\left|\langle n_6|{\textstyle{\frac{1}{2}}}(-i\beta_1-\beta_2e^{i\theta})\rangle\right|^2.
\end{eqnarray}
These sums have been previously computed
\cite{Freyberger93,Braunstein90,Vogel93} and we find
\begin{eqnarray}
K_{43}&=&
e^{-\frac{1}{2}(|\beta_1|^2+|\beta_2|^2)}
\left|\frac{\beta_1+\beta_2 e^{i\theta}}{\beta_1-\beta_2 e^{i\theta}}\right|^{n_{43}}
I_{|n_{43}|}\left({\textstyle{\frac{1}{2}}}|\beta_1^2-\beta_2^2 e^{2i\theta}|\right),
\label{eq:K43}\\
\noalign{\hbox{\rm and}}
K_{65}&=&
e^{-\frac{1}{2}(|\beta_1|^2+|\beta_2|^2)}
\left|\frac{i\beta_1+\beta_2 e^{i\theta}}{i\beta_1-\beta_2 e^{i\theta}}\right|^{n_{65}}
I_{|n_{65}|}\left({\textstyle{\frac{1}{2}}}|\beta_1^2+\beta_2^2 e^{2i\theta}|\right),
\label{eq:K65}
\end{eqnarray}
where $I_\nu(z)$ denotes the modified Bessel function of order $\nu$.

%%%%%%%%%%%%%%%%%%%%%%%%%%%%%%%%%%%%%%%%%%%%%%%%%%%%%%%%%%%%%%%%%%%%%%%%%%%%
\subsection{Phase distributions from photon counts}
\label{sec:distributions}
%%%%%%%%%%%%%%%%%%%%%%%%%%%%%%%%%%%%%%%%%%%%%%%%%%%%%%%%%%%%%%%%%%%%%%%%%%%%

An eight-port homodyne detector as shown in
Fig.~\ref{fig:direct} measures the photon number differences $n_{43}$ and
$n_{65}$ at the output.  What we obtain are the
probabilities, $W(n_{43},n_{65})$, as defined in Eq.
(\ref{eq:joint-def}), and we would like to use them to construct a phase
distribution following the prescription given by NFM and summarized in
Eqs.  (\ref{eq:NFM-cos}) and (\ref{eq:NFM-sin}).  This can be done by
representing the pairs $(n_{43},n_{65})$ as points in a two-dimensional
space where $n_{43}$ is the $x$ coordinate and $n_{65}$ is the $y$
coordinate.  Eqs.  (\ref{eq:NFM-cos}) and (\ref{eq:NFM-sin}) imply that
a difference phase of $\varphi$ corresponds to a ray in this space which
starts at the origin and makes an angle $\varphi$ with the positive
$x$-axis.  The probability assigned to $\varphi$ is just the sum of the
probabilities of the points, $(n_{43},n_{65})$, which the ray passes
through.

There are two problems with this prescription.  The first is that the
distribution will not be smooth; it will consist of a number of spikes.
Certain rays will intersect no $(n_{43},n_{65})$ points yielding a value
of zero for the probability of the corresponding angle, whereas a nearby
ray will hit such a point which can lead to a nonzero probability for
its angle.  What is needed is a way to smooth the distribution and this
has been provided by NFM, and we shall discuss it shortly. The second
problem has to do with the contribution of the origin, $n_{43}=0$ and
$n_{65}=0$, to the phase distribution.  A glance at Eqs.
(\ref{eq:NFM-cos}) and (\ref{eq:NFM-sin}) show that the sine and cosine
are not defined at this point which implies that the angle is not
defined either.  One now has to decide what to do with the probability
corresponding to this point when constructing the phase distribution.
NFM throw out the data corresponding to this point and renormalize the
remaining probabilities, $W(n_{43},n_{65})$ for $n _{43},n_{65}\neq0$,
accordingly.  Another possibility is to associate $W(0,0)$ with a
uniform distribution of the angle.  The philosophy behind this approach
is that because we have no information about how to assign $W(0,0)$ to any
angle, we apportion it equally among all angles.

We follow the prescription in Sec.~\ref{sec:operators} to obtain the two
possible phase difference distributions
\begin{eqnarray}
{\cal P}_0(\varphi)&=&\langle\delta(\varphi-\hat\varphi_M)\rangle=
        \left\langle\delta\left[\varphi-\arctan\left(\frac{\hat S_M}{\hat C_M}
                \right)\right]\right\rangle\nonumber\\
&=&\frac{1}{1-W(0,0)}\sum_{n_{43},n_{65}\neq0}W(n_{43},n_{65})
        \delta\left(\varphi-\arctan\frac{n_{65}}{n_{43}}\right),
\label{eq:P0}
\end{eqnarray}
and
\begin{equation}
{\cal P}_1(\varphi)=\frac{1}{2\pi}W(0,0)+\sum_{n_{43},n_{65}\neq0}W(n_{43},n_{65})
        \delta\left(\varphi-\arctan\frac{n_{65}}{n_{43}}\right),
\label{eq:P1}
\end{equation}
where ${\cal P}_0(\varphi)$ is the distribution in which the origin has been
eliminated and ${\cal P}_1(\varphi)$ is the distribution in which it has been
included. In the above expressions [Eqs.~(\ref{eq:P0}) and (\ref{eq:P1})] and
in all that follow one should be careful to choose the right branch of the
function $\arctan$ according to the signs of $n_{43}$ and $n_{65}$.

The averaging procedure which was developed by NFM to smooth the phase
distribution, and which is also used by Torgerson and Mandel, works in
the following way.  A phase shift of $\theta$ is introduced into the
field entering port 2 of the 8-port detector.  This produces new
probabilities for the photon number differences at the output which we
shall denote by $W(n_{43},n_{65}| e^{i\theta})$.  Going back to our
two-dimensional space, because the phase shift maps a phase difference
of $\varphi$ into one of $\varphi-\theta$, a ray which makes an angle of
$\varphi-\theta$ with the $x$-axis is associated with a phase shift of
$\varphi$.  This gives us a difference-phase distribution for each value
of the phase shift $\theta$.
\begin{equation}
{\cal P}_0(\varphi|\theta)=\langle\delta[\varphi-(\hat\varphi_M-\theta)]\rangle={\cal N}_0
\sum_{n_{43},n_{65}\neq0}W(n_{43},n_{65}| e^{i\theta})
\delta\left[\varphi-\left(\arctan\frac{n_{65}}{n_{43}}-\theta\right)\right],
\label{eq:P0-theta}
\end{equation}
and
\begin{equation}
{\cal P}_1(\varphi|\theta)=\frac{1}{2\pi}W(0,0| e^{i\theta})+
\sum_{n_{43},n_{65}\neq0}W(n_{43},n_{65}| e^{i\theta})
\delta\left[\varphi-\left(\arctan\frac{n_{65}}{n_{43}}-\theta\right)\right].
\label{eq:P1-theta}
\end{equation}
We then obtain a final phase distribution
by averaging
\begin{equation}
\overline{\cal P}(\varphi)=\frac{1}{2\pi}\int_0^{2\pi}d\theta\,
{\cal P}(\varphi|\theta)
\label{eq:P-averaged}
\end{equation}
over all of them. The $\theta$ integrations are trivially performed, giving us
\begin{equation}
\overline{{\cal P}_0}(\varphi)=\frac{{\cal N}_0}{2\pi}
\sum_{n_{43},n_{65}\neq0}W\left(n_{43},n_{65}|
e^{i(\arctan n_{65}/n_{43}-\varphi)}\right),
\label{eq:P0-general}
\end{equation}
and
\begin{equation}
\overline{{\cal P}_1}(\varphi)=\frac{1}{2\pi}\left\{
\frac{1}{2\pi}\int_0^{2\pi}d\theta\, W(0,0| e^{i\theta})+
\sum_{n_{43},n_{65}\neq0}W\left(n_{43},n_{65}|
e^{i(\arctan n_{65}/n_{43}-\varphi)}\right)\right\},
\label{eq:P1-general}
\end{equation}
where, as before, $\overline{{\cal P}_0}(\varphi)$ does not include the point
$n_{43}=n_{65}=0$ and $\overline{{\cal P}_1}(\varphi)$ does.  Note that in calculating
$\overline{{\cal P}_0}(\varphi)$ we have first averaged the unnormalized distributions and
then normalized the result.  The constant ${\cal N}_0$ in Eq.
(\ref{eq:P0-theta}) is chosen so that
$\int_0^{2\pi}d\varphi \overline{{\cal P}_0}(\varphi)=1$.  This is the procedure adopted
by NFM.  However, one could just as well normalize the distribution for
each value of $\theta$ and then average the result.  This will not give
the same result as the first procedure because the normalization
constant for each value of $\theta$, $(1-W(0,0| e^{i\theta}))^{-1}$,
depends on $\theta$.  There is no obvious reason to prefer one procedure
over the other, so we simply use the one which was chosen by NFM.

%%%%%%%%%%%%%%%%%%%%%%%%%%%%%%%%%%%%%%%%%%%%%%%%%%%%%%%%%%%%%%%%%%%%%%%%%%%%
\section{Schemes}
\label{sec:schemes}
%%%%%%%%%%%%%%%%%%%%%%%%%%%%%%%%%%%%%%%%%%%%%%%%%%%%%%%%%%%%%%%%%%%%%%%%%%%%

Using the eight-port homodyne interferometer described in
Sec.~\ref{sec:8-port}, Torgerson and Mandel~\cite{Torgerson96} measure
the phase difference between two coherent optical fields by (a) beating
the two fields against each other and (b) by beating them against a
common local oscillator.  Torgerson and Mandel refer to these as the
direct and indirect measurements.  In this section we briefly describe
these schemes and derive analytical expressions for the
corresponding phase distributions.

%%%%%%%%%%%%%%%%%%%%%%%%%%%%%%%%%%%%%%%%%%%%%%%%%%%%%%%%%%%%%%%%%%%%%%%%%%%%
\subsection{Direct scheme}
\label{sec:direct}
%%%%%%%%%%%%%%%%%%%%%%%%%%%%%%%%%%%%%%%%%%%%%%%%%%%%%%%%%%%%%%%%%%%%%%%%%%%%

The direct measurement is made by beating the two input fields against
each other in a single eight-port homodyne interferometer.  The
expression for the phase distribution is then simply
Eq.~(\ref{eq:P0-general}) [or Eq.~(\ref{eq:P1-general})].  In what
follows, we evaluate this expression in the limits of strong
and weak fields.

%%%%%%%%%%%%%%%%%%%%%%%%%%%%%%%%%%%%%%%%%%%%%%%%%%%%%%%%%%%%%%%%%%%%%%%%%%%%%%
\subsubsection{Strong field limit}
\label{sec:direct-strong}
%%%%%%%%%%%%%%%%%%%%%%%%%%%%%%%%%%%%%%%%%%%%%%%%%%%%%%%%%%%%%%%%%%%%%%%%%%%%%%

When one of the fields, say field 2, is strong we can use the known
result for the strong
local oscillator limit~\cite{Freyberger93a}, where field 2 plays the
role of the local oscillator.  In this case, the phase distribution
(both ${\cal P}_0$ and ${\cal P}_1$) reduces to
\begin{eqnarray}
{\cal P}(\varphi)&=&\frac{1}{\pi}\int d^2\beta_1\int d^2\beta_2
\int_0^\infty r dr P(\beta_1,\beta_2)
\nonumber\\
&&{}\times\exp\Bigl\{
-\bigl[r\cos\varphi-|\beta_1|\cos(\phi_2-\phi_1)\bigr]^2
-\bigl[r\sin\varphi-|\beta_1|\sin(\phi_2-\phi_1)\bigr]^2\Bigr\},
\label{eq:direct-strong}
\end{eqnarray}
where $\beta_j=|\beta_j|e^{i\phi_j}$, for $j=1,2$.

%%%%%%%%%%%%%%%%%%%%%%%%%%%%%%%%%%%%%%%%%%%%%%%%%%%%%%%%%%%%%%%%%%%%%%%%%%%%%%%%
\subsubsection{Weak fields limit}
\label{sec:direct-weak}
%%%%%%%%%%%%%%%%%%%%%%%%%%%%%%%%%%%%%%%%%%%%%%%%%%%%%%%%%%%%%%%%%%%%%%%%%%%%%%%%

When both fields are weak, we use the relation
\[I_\nu(z)\simeq\frac{1}{\nu!}\left(\frac{z}{2}\right)^\nu\]
for the modified Bessel function, which allows us to write the kernels in
Eq.~(\ref{eq:joint}) as
\begin{equation}
K_{43}\simeq\left\{
\begin{array}{lr}
\displaystyle{\frac{e^{-\frac{1}{2}(|\beta_1|^2+|\beta_2|^2)}}{4^{n_{43}} n_{43}!}}
\left|\beta_1+\beta_2e^{i\theta}\right|^{2n_{43}},& n_{43}>0,\\
e^{-\frac{1}{2}(|\beta_1|^2+|\beta_2|^2)},& n_{43}=0,\\
\displaystyle{\frac{e^{-\frac{1}{2}(|\beta_1|^2+|\beta_2|^2)}}{4^{|n_{43}|}|n_{43}|!}}
\left|\beta_1-\beta_2e^{i\theta}\right|^{2|n_{43}|},& n_{43}<0,
\end{array}
\right.
\end{equation}
\begin{equation}
K_{65}\simeq\left\{
\begin{array}{lr}
\displaystyle{\frac{e^{-\frac{1}{2}(|\beta_1|^2+|\beta_2|^2)}}{4^{n_{65}} n_{65}!}}
\left|i\beta_1+\beta_2e^{i\theta}\right|^{2n_{65}},& n_{65}>0,\\
e^{-\frac{1}{2}(|\beta_1|^2+|\beta_2|^2)},& n_{65}=0,\\
\displaystyle{\frac{e^{-\frac{1}{2}(|\beta_1|^2+|\beta_2|^2)}}{4^{|n_{65}|}|n_{65}|!}}
\left|i\beta_1-\beta_2e^{i\theta}\right|^{2|n_{65}|},& n_{65}<0.
\end{array}
\right.
\end{equation}

The remaining integral in Eq.~(\ref{eq:P1-general}) is now trivially performed
yielding
\[\frac{1}{2\pi}\int_0^{2\pi}d\theta\, W(0,0| e^{i\theta})=\int d^2\beta_1 d^2\beta_2
P(\beta_1,\beta_2)e^{-(|\beta_1|^2+|\beta_2|^2)}.\]
In this limit, the sums in Eqs.~(\ref{eq:P0-general}) and (\ref{eq:P1-general})
have only
contributions from those terms in the lowest order in the coherent
state amplitudes $|\beta_j|$.  Hence only terms with $n_{43}=-1,0,1$
and $n_{65}=-1,0,1$ contribute and we arrive finally at
\begin{equation}
\overline{{\cal P}_0}(\varphi)\simeq\frac{{\cal N}_0}{2\pi}\int d^2\beta_1
d^2\beta_2 P(\beta_1,\beta_2)[|\beta-1|^2+|\beta_2|^2+
2|\beta_1||\beta_2|\cos(\varphi+\phi_1-\phi_2)],
\label{eq:direct-Mandel.gen}
\end{equation}
and
\begin{equation}
\overline{{\cal P}_1}(\varphi)\simeq\frac{1}{2\pi}\int d^2\beta_1 d^2\beta_2
P(\beta_1,\beta_2)[1+2|\beta_1||\beta_2|
        \cos(\varphi+\phi_1-\phi_2)],
\label{eq:direct-with.gen}
\end{equation}
up to corrections which are quartic in $|\beta_j|$.  Note that Eq.
(\ref{eq:direct-Mandel.gen}) reproduces exactly the result obtained by
Torgerson and Mandel~\cite{Torgerson96},
\begin{equation}
\overline{{\cal P}_0}(\varphi)\simeq\frac{1}{2\pi}
        \left(1+\frac{2\cos(\varphi+\phi_1-\phi_2)}
        {|\beta_1|/|\beta_2|+|\beta_2|/|\beta_1|}\right),
\label{eq:direct-Mandel}
\end{equation}
for input fields in coherent states $|\beta_1\rangle$ and $|\beta_2\rangle$.
In this case, the phase distribution $\overline{{\cal P}_1}(\varphi)$, which
includes the point $n_{43}=n_{65}=0$, reduces to
\begin{equation}
\overline{{\cal P}_1}(\varphi)\simeq\frac{1}{2\pi}\Bigl[1+2|\beta_1||\beta_2|
        \cos(\varphi+\phi_1-\phi_2)\Bigr].
\label{eq:direct-with}
\end{equation}
%

%%%%%%%%%%%%%%%%%%%%%%%%%%%%%%%%%%%%%%%%%%%%%%%%%%%%%%%%%%%%%%%%%%%%%%%%%%%%%%%%%
\subsection{Indirect scheme}
\label{sec:indirect}
%%%%%%%%%%%%%%%%%%%%%%%%%%%%%%%%%%%%%%%%%%%%%%%%%%%%%%%%%%%%%%%%%%%%%%%%%%%%%%%%%

\begin{figure}
\centerline{\epsfig{file=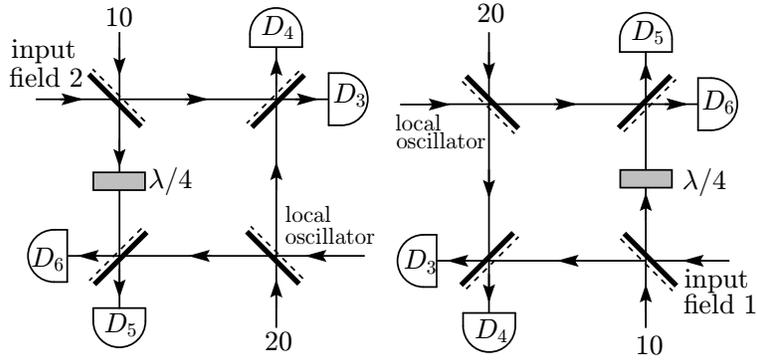,width=4in}}\vspace{1ex}
\caption{Indirect measurement scheme:  A pair of eight-port homodyne
detectors is used to measure the phase difference between the input
fields 1 and 2.  Each field is beat against a common
strong local oscillator with coherent amplitude $\alpha$.
Each eight-port detector has four
detectors measuring the photon count differences $n_{43}$ and $n_{65}$.
Two phase distributions are obtained and combined to give the phase
distribution for the phase difference.}
\label{fig:indirect}
\end{figure}

The indirect measurement scheme consists of two eight-port homodyne
detectors, each performing a measurement of the phase distribution for one
of the fields relative to a common strong local oscillator, $\alpha$, as
seen in Fig.~\ref{fig:indirect}.  Making no assumptions about the incoming
state we could define the relative-phase distribution as the convolution
\begin{equation}
{\cal P}(\varphi)=\int_0^{2\pi}d\varphi_1 {\cal P}(\varphi_1,
\varphi_1-\varphi),
\label{eq:convolution}
\end{equation}
of the joint phase distribution
\begin{equation}
{\cal P}(\varphi_1,\varphi_2) \equiv \frac{1}{\pi} \int_0^\infty r_1 dr_1
\int_0^\infty r_2 dr_2 \langle r_1e^{i\varphi_1}, r_2e^{i\varphi_2} |
\hat\rho | r_1e^{i\varphi_1}, r_2e^{i\varphi_2} \rangle
\label{eq:integrated-Q}
\end{equation}
which is the radially integrated two-mode $Q$-function,
in terms of the two-particle state $\hat\rho$. We note that this
reduces to the expected joint phase distribution
\begin{equation}
{\cal P}(\varphi_1,\varphi_2)={\cal P}_1(\varphi_1){\cal P}_2(\varphi_2).
\end{equation}
when the input fields are in the product state
$\hat\rho = \hat\rho_1\otimes\hat\rho_2$ and where the individual
phase distributions ${\cal P}_i(\varphi_i)$ have their obvious meaning.

Let us now restrict our attention to the case where the input fields
1 and 2 have such a factorable state, so the two-mode $P$-function can be
written $P(\beta_1,\beta_2)=P(\beta_1)P(\beta_2)$. Each phase distribution
now takes the form
\begin{eqnarray}
{\cal P}_i(\varphi_i)&=&\frac{1}{\pi}\int d^2\beta_i P(\beta_i)\int_0^\infty r_i
dr_i\nonumber\\
&&{}\times\exp\Bigl\{
-\bigl[r_i\cos\varphi_i-|\beta_i|\cos(\phi_0-\phi_i)\bigr]^2
-\bigl[r_i\sin\varphi_i-|\beta_i|\sin(\phi_0-\phi_i)\bigr]^2
\Bigr\},
\end{eqnarray}
where $\phi_0$ is the phase of the local oscillator and the $P(\beta_i),
i=1,2$ are Glauber-Sudarshan distributions for the fields 1 and 2.
Evaluating the integral in Eq.~(\ref{eq:convolution}) we obtain the
expression
\begin{eqnarray}
{\cal P}(\varphi)&=&\frac{2}{\pi}\int d^2\beta_1P(\beta_1)\int d^2\beta_2 \
P(\beta_2) e^{-|\beta_1|^2-|\beta_2|^2}\nonumber\\
&&{}\times\int_0^\infty r_1
dr_1\int_0^\infty r_2 dr_2\; e^{-r_1^2-r_2^2}
I_0\bigl(2|r_1\beta_1+r_2\beta_2e^{-i\varphi}|\bigr)
\label{eq:indirect-general}
\end{eqnarray}
for the phase distribution of the indirect measurement. Again we now consider
the limiting cases of this exact expression.

%%%%%%%%%%%%%%%%%%%%%%%%%%%%%%%%%%%%%%%%%%%%%%%%%%%%%%%%%%%%%%%%%%%%%%%%%%%%%%%%
\subsubsection{Strong field limit}
\label{sec:indirect-strong}
%%%%%%%%%%%%%%%%%%%%%%%%%%%%%%%%%%%%%%%%%%%%%%%%%%%%%%%%%%%%%%%%%%%%%%%%%%%%%%%%

In the case when the input field 2 is strong but still much weaker
than the local oscillator, Eq.~(\ref{eq:indirect-general})
approaches the result Eq.~(\ref{eq:direct-strong}) for the direct
measurement very quickly.  This behavior can be seen from
Fig.~\ref{fig:distribution}, where we have plotted both expressions
for fields 1 and 2 in coherent states and $|\beta_2/\beta_1|=4$.
For higher values of $|\beta_2|$ the two curves coincide.

%%%%%%%%%%%%%%%%%%%%%%%%%%%%%%%%%%%%%%%%%%%%%%%%%%%%%%%%%%%%%%%%%%%%%%%%%%%%%%%%
\subsubsection{Weak fields limit}
\label{sec:indirect-weak}
%%%%%%%%%%%%%%%%%%%%%%%%%%%%%%%%%%%%%%%%%%%%%%%%%%%%%%%%%%%%%%%%%%%%%%%%%%%%%%%%

In the case when both fields are weak, the Bessel function in
Eq.~(\ref{eq:indirect-general}) can be written as
\[I_0(|z|)\simeq 1 + \frac{|z|^2}{4}\]
and the integral is easily performed yielding
\begin{equation}
{\cal P}(\varphi)\simeq\frac{1}{2\pi}\int d^2\beta_1 d^2\beta_2 P(\beta_1,\beta_2)
        \left[1+\frac{\pi}{2}|\beta_1||\beta_2|
        \cos(\varphi+\phi_1-\phi_2)\right]
\end{equation}
to lowest order for the phase distribution.

We conclude this section by noting that, in contrast to the direct
measurement, in this indirect measurement the contribution from
ambiguous data is always negligible due to the presence of the strong
local oscillator.

%%%%%%%%%%%%%%%%%%%%%%%%%%%%%%%%%%%%%%%%%%%%%%%%%%%%%%%%%%%%%%%%%%%%%%%%%%%%
\section{States with identical phase distributions}
\label{sec:identical}
%%%%%%%%%%%%%%%%%%%%%%%%%%%%%%%%%%%%%%%%%%%%%%%%%%%%%%%%%%%%%%%%%%%%%%%%%%%%

An unusual feature of the direct phase measurement scheme of
Torgerson and Mandel (TM) is that any pair of two-mode input states
which differ in their vacuum component, $|0,0\rangle$, will have
the same phase distribution \cite{Noh92,Noh92a}. For states with
large photon numbers this is of little importance, but for states
with very small photon numbers it has interesting consequences.

Let us first briefly show that this is true.  Let
$p(n_3,n_4,n_5,n_6|\theta)$ be
the probability that $D_3$ detects $n_3$ photons, $D_4$ detects $n_4$
photons, etc. (see Fig. \ref{fig:direct}),
given that the phase of the input in port 2 has been
shifted by $\theta$.  If the two-field input state is $|\psi_{\rm
in}\rangle$ and $\hat U(\theta)$ describes the action of the entire
apparatus, with the phase shifter in port 2 set to $\theta$, then
\begin{equation}
p(n_3,n_4,n_5,n_6|\theta)=|\langle n_3,n_4,n_5,n_6|\hat U(\theta)|
\psi_{\rm in},0,0\rangle|^2,
\label{eq:interf}
\end{equation}
where the two zeros appended to $|\psi_{\rm in}\rangle$ indicate that
the other two inputs are in the vacuum state.  Let us express
$|\psi_{\rm in}\rangle$ as
\begin{equation}
|\psi_{\rm in}\rangle=c_0|0,0\rangle+|\Phi\rangle,
\label{eq:vacuum-sep1}
\end{equation}
where $\langle\Phi|0,0\rangle=0$ and $|c_0|^2+\|\Phi\|^2=1$.
Because the beam splitters and the phase shifter conserve total photon
number we have
\begin{eqnarray}
p(0,0,0,0|\theta)&=&|c_0|^2, \nonumber\\
p(n_3,n_4,n_5,n_6|\theta)&=&|\langle n_3,n_4,n_5,n_6|\hat U(\theta)|\Phi,0,0\rangle|^2,
\quad \hbox{\rm if $n_3+n_4+n_5+n_6\neq 0$}.
\label{eq:vacuum-sep2}
\end{eqnarray}
Now consider the input state
\begin{equation}
|\psi'_{\rm in}\rangle=\frac{1}{\|\Phi\|}|\Phi\rangle
\end{equation}
and denote the output photon number probabilities as
$p'(n_3,n_4,n_5,n_6|\theta)$.  (We note that this vacuum-depleted state
$|\psi'_{\rm in}\rangle$ will be typically an entangled state when the
original $|\psi_{\rm in}\rangle$ is a two-mode product state.)
Eq.~(\ref{eq:vacuum-sep2}) implies that for $n_3+n_4+n_5+n_6>0$,
\begin{equation}
p'(n_3,n_4,n_5,n_6|\theta)=\frac{1}{\|\Phi\|^2}p(n_3,n_4,n_5,n_6|\theta).
\label{eq:implies}
\end{equation}
Because events corresponding to $n_j=0$, $j=1,\dots,4$ are discarded, the
probabilities $p(n_3,n_4,n_5,n_6|\theta)$ and
$p'(n_3,n_4,n_5,n_6|\theta)$, for $n_3+n_4+n_5+n_6>0$, determine the
TM phase distributions for $|\psi_{\rm in}\rangle$ and $|\psi'_{\rm
in}\rangle$, respectively, and they are the same up to an overall
factor.  When the TM phase distributions for the two states are computed
and normalized this factor disappears.  Therefore, $|\psi_{\rm
in}\rangle$ and $|\psi'_{\rm in}\rangle$ have the same TM phase
distribution.

Let us see what this implies for the input state considered by Torgerson
and Mandel, two weak coherent states.  We have for
$|\beta_1|,|\beta_2|\ll 1$
\begin{equation}
|\psi_{\rm in}\rangle=|\beta_1\rangle\otimes|\beta_2\rangle\simeq
e^{-\frac{1}{2}(|\beta_1|^2+|\beta_2|^2)}
(|0,0\rangle+\beta_1|1,0\rangle+\beta_2|0,1\rangle).
\label{eq:psi-in}
\end{equation}

A consequence of the result in the previous paragraph is that the state
\begin{equation}
|\psi'_{\rm in}\rangle=\left(\frac{1}{|\beta_1|^2+|\beta_2|^2}\right)^{1/2}
(\beta_1|1,0\rangle+\beta_2|0,1\rangle)
\label{eq:psi'-in}
\end{equation}
has the same TM phase distribution as $|\psi_{\rm in}\rangle$ which is given by
Eq.~(\ref{eq:direct-Mandel}), i.e.
\[
{\cal P}(\varphi)\simeq\frac{1}{2\pi}\left(1+\frac{2\cos(\varphi+\phi_1-\phi_2)}
        {|\beta_1|/|\beta_2|+|\beta_2|/|\beta_1|}\right).
\]
This fact has been noted by NFM \cite{Noh92,Noh92a}. Torgerson and
Mandel pointed out that the agreement between the above phase
distribution, which agreed with their measurements, and the London
distribution for $|\psi_{\rm in}\rangle$ is poor. Suppose we
instead compare Eq.~(\ref{eq:direct-Mandel}) to the London
distribution for the state $|\psi'_{\rm in}\rangle$.  We find that
they are identical.

One interpretation of these results is the following.  If ${\cal
P}(\varphi)$ is taken to be the TM phase distribution of
$|\psi_{\rm in}\rangle$ one has the rather puzzling result that as
$|\beta_1|$ and $|\beta_2|$ go to zero, with $|\beta_1|/|\beta_2|$
fixed, the phase distribution does not change.  One would expect
that as $|\psi_{\rm in}\rangle$ becomes closer to the vacuum its
phase distribution would become more uniform.  On the other hand,
if ${\cal P}(\varphi)$ is interpreted as the phase distribution for
$|\psi'_{\rm in}\rangle$ the puzzle disappears.  If $|\beta_1|$ and
$|\beta_2|$ go to zero, with $|\beta_1|/|\beta_2|$ fixed, the state
$|\psi'_{\rm in}\rangle$ does not change and, therefore, neither
should its phase distribution.  Furthermore, its operational phase
distribution (obtained via the direct measurement procedure) agrees
with its London phase distribution.  Because it discards data, the
measurement procedure of Torgerson and Mandel is insensitive to the
vacuum component of input states, which, along with the previous
two observations, suggests that what is being measured is the phase
distribution of $|\psi'_{\rm in}\rangle$ and not that of
$|\psi_{\rm in}\rangle$.  This observation is closely related to
one by Hradil in his discussion of the original Noh, Foug\`eres,
and Mandel proposal~\cite{Noh91,Noh92,Hradil93}.

Another objection which can be raised to the interpretation of
Eq.~(\ref{eq:direct-Mandel}) as a phase distribution for
$|\psi_{\rm in}\rangle$ has to do with the connection between a
phase distribution and the rotational properties of its
corresponding state in phase space. A useful property for a phase
distribution to have is providing information about how
distinguishable a state is from a rotated version of itself.  A
phase distribution with this property will be able to tell us how
useful a given state is for the measurement of phase shifts.  In
the case of two-mode states we would like the difference-phase
distribution of a state $|\psi\rangle$ to give us information about
$|\langle\psi|\hat U_D(\varphi_0)|\psi\rangle|$, which is a measure
of how distinguishable $|\psi\rangle$ is from its difference-phase
rotated version, $\hat U_D(\varphi_0)|\psi\rangle$.  Here the
operator $\hat U_D(\varphi_0)=\exp[i\frac{\varphi_0}{2}(\hat
n_1-\hat n_2)]$ rotates mode 1 by $\varphi_0/2$ and mode 2 by
$-\varphi_0/2$, thereby changing the phase difference between the
modes by $\varphi_0$.  For $|\psi_{\rm in}\rangle$ we find
\begin{equation}
|\langle\psi_{\rm in}|\hat U_D(\varphi_0)|\psi_{\rm in}\rangle|\simeq
e^{-(|\beta_1|^2+|\beta_2|^2)}
\left|1+|\beta_1|^2e^{i\varphi_0/2}+|\beta_2|^2e^{-i\varphi_0/2}\right|,
\label{eq:rotation}
\end{equation}
which goes to one as $|\beta_1|$ and $|\beta_2|$ go to zero.  This shows
that $|\psi_{\rm in}\rangle$ and its rotated version become
indistinguishable in this limit and $|\psi_{\rm in}\rangle$ would be of
little use in detecting a phase shift.  This is contrary to what one
sees from the TM phase distribution in
Eq.~(\ref{eq:direct-Mandel}).  If $|\beta_1|=|\beta_2|$ the TM phase
distributions ${\cal P}(\varphi)$ and ${\cal P}(\varphi+\varphi_0)$ are clearly
distinguishable if $\varphi_0$ is, for example, $\pi/4$ while according
to Eq.~(\ref{eq:rotation}) (if $|\beta_1|,|\beta_2|\ll 1$) the states
$|\psi_{\rm in}\rangle$ and $\hat U_D(\varphi_0)|\psi_{\rm in}\rangle$ are
not.  On the other hand, we find
\begin{equation}
\langle\psi'_{\rm in}|\hat U_D(\varphi_0)|\psi'_{\rm in}\rangle|=
\frac{1}{|\beta_1|/|\beta_2|+|\beta_2|/|\beta_1|}
\left[\frac{|\beta_1|^2}{|\beta_2|^2}+
\frac{|\beta_2|^2}{|\beta_1|^2}+2\cos\varphi_0\right]^{1/2},
\label{eq:rotation'}
\end{equation}
which is independent of the size of $|\beta_1|$ and $|\beta_2|$ and
depends only on their ratio.  Eq.~(\ref{eq:rotation'}) implies that if
$|\beta_1|=|\beta_2|$ and $\varphi_0=\pi/4$, then $|\psi'_{\rm
in}\rangle$ and its rotated version can be distinguished which is
consistent with what we expect from its London distribution.  This again
suggests interpreting the TM phase distribution in
Eq.~(\ref{eq:direct-Mandel}) as being associated with $|\psi'_{\rm
in}\rangle$ rather than $|\psi_{\rm in}\rangle$.

It should be noted that if all of the data is kept these questions of
interpretation do not arise.  In that case the vacuum state does
contribute to the phase distribution and the phase distributions for
$|\psi_{\rm in}\rangle$ and $|\psi'_{\rm in}\rangle$ are no longer the
same.

%%%%%%%%%%%%%%%%%%%%%%%%%%%%%%%%%%%%%%%%%%%%%%%%%%%%%%%%%%%%%%%%%%%%%%%%%%%%
\section{To post-select or not?}
\label{sec:post-select}
%%%%%%%%%%%%%%%%%%%%%%%%%%%%%%%%%%%%%%%%%%%%%%%%%%%%%%%%%%%%%%%%%%%%%%%%%%%%

In Section~\ref{sec:direct} we have discussed two possibilities of data
analysis and have shown the phase distributions corresponding to
retaining or discarding ambiguous data.  In this section we briefly
return to this point.

The data analysis used in Eqs.~(\ref{eq:P0}) and (\ref{eq:P1}) may
be compactly
summarized by considering the distribution for the random variable
$Z=f(X,Y)$ given the distribution ${\cal P}(X,Y)$ for the random variables $X$
and $Y$.  This can be written as
\[{\cal P}(Z) \propto \int dX dY \delta(Z-f(X,Y)) {\cal P}(X,Y).\]
When the data, which in this case is a pair $(X, Y)$, unambiguously
determines $Z$ the above formula is trivial; however, when the data
leads to no single $Z$ then we must make the assignment based on other
considerations.  We suggest spreading the probability evenly amongst the
consistent values---uniformly across the $2\pi$ radians.  This is
consistent with a maximum entropy assignment given no other information
about the incoming state.

In contrast the approach used in Refs.
\cite{Noh91,Noh92,Noh92a,Noh93a} discards data obtained for
$n_{43}=n_{65}=0$.  The excluded data does not appear to be very
useful for informing us about phase. Thus, it is justifiable to
select only the unambiguous data as has been done in Ref.
\cite{Noh91,Noh92,Noh92a,Noh93a}. However, a comparison of the
behavior of the direct scheme without and with postselection for
weak fields, Eqs.~(\ref{eq:direct-with})
and~(\ref{eq:direct-Mandel}) respectively, shows that the latter
has a significantly narrower distribution.  We have argued in the
previous section that one resolution of this apparent paradox is
that by discarding data we are changing the state which we measure.
Another way of looking at this issue is in terms of the sensitivity
of the measurement.

It seems that when data is discarded we achieve a greater measurement
sensitivity, because the resulting phase distribution is narrower than
would be the case if all the data were kept.  How can we reconcile this
with Shannon's information theory which teaches us that we cannot
improve sensitivity by discarding information?  It must be that the
apparent difference in widths---and naively sensitivity---for these two
distributions is in some sense illusory.  The resolution to this
`paradox' is that the discarded data carries information about the
overall resources used which would need to be factored into any
meaningful measure of sensitivity.  The deeper question of how we may
compare phase distributions between different schemes, with respect to
the cost of resources involved, is beyond the scope of this paper.
However, it is clear that for states of very small photon number most
measurements will yield a null result and will have to be thrown out.
Therefore, a large number of measurements will have to be made in order
to obtain only a few usable ones, which suggests a large expenditure of
resources, and a corresponding decrease of sensitivity.

%%%%%%%%%%%%%%%%%%%%%%%%%%%%%%%%%%%%%%%%%%%%%%%%%%%%%%%%%%%%%%%%%%%%%%%%%%%%
\section{Conclusion}
\label{sec:conclusion}
%%%%%%%%%%%%%%%%%%%%%%%%%%%%%%%%%%%%%%%%%%%%%%%%%%%%%%%%%%%%%%%%%%%%%%%%%%%%

We may generate the distribution for the relative phases of a pair of
coherent states $\beta_1$ and $\beta_2$ using the direct or indirect
schemes described here.  Our results confirm Torgerson and Mandel's
claim of the non-uniqueness of operational phase operators; the two
measurement procedures lead to different phase distributions.

\begin{figure}
\centerline{\epsfig{file=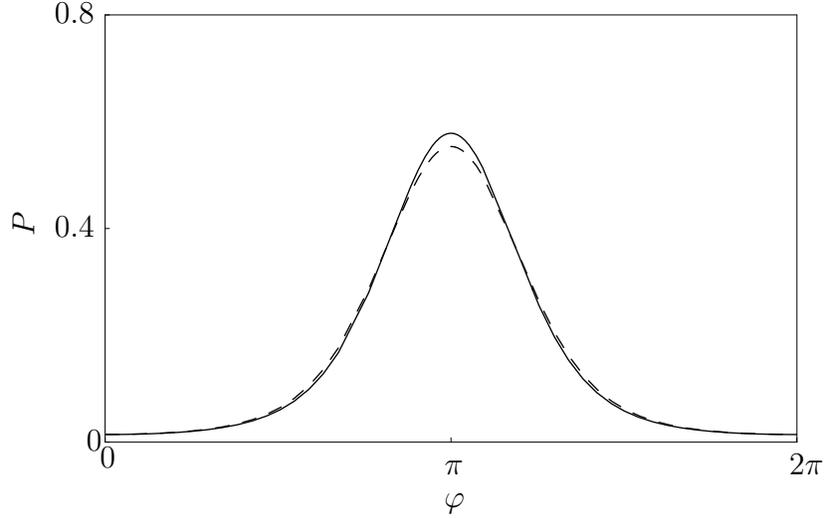}}\vspace{1ex}
\caption{Comparison between phase distributions for both direct (solid
line) and indirect measurement (dashed line), in the strong field limit.
In this limit, the phase distribution of the direct scheme follows from
Eq.~(\protect\ref{eq:direct-strong}), whereas for the indirect
measurement, we have used the general expression
Eq.~(\protect\ref{eq:indirect-general}).  For both curves we have chosen
the input fields 1 and 2 in coherent states with $\beta_1=1$ and $\beta_2=-4$.}
\label{fig:distribution}
\end{figure}

A real comparison between the schemes' sensitivity must await a
careful analysis of the resources used.  Notwithstanding this, we
shall make a zeroth-order comparison here in the absence of
postselection. As we noted already above
(Section~\ref{sec:indirect-strong}) in the strong field limit
($1\ll|\beta_2|\ll\alpha$ and $|\beta_1|\ll|\beta_2|$) the two
schemes, direct and indirect, go to the same distribution (see
Fig.~\ref{fig:distribution}). In the weak fields limit we have
calculated, however, the situation is somewhat more interesting:
The direct scheme reproduces the statistics of the London
distribution (also called Suskind-Glogower or Pegg-Barnett) applied
to the difference-phase \cite{Torgerson96}.  By contrast, the
indirect scheme produces a phase distribution which is slightly
broader by the ratio of $4/\pi$ ($\simeq 1.27$).  These facts
surprised the authors:  Both that one scheme is a realization of
the London distribution in the limit where both input fields are
weak coherent states, and that the other scheme does not perform
identically.  Clearly, interesting questions remain.

%%%%%%%%%%%%%%%%%%%%%%%%%%%%%%%%%%%%%%%%%%%%%%%%%%%%%%%%%%%%%%%%%%%%%%%%%%%%
\acknowledgments
%%%%%%%%%%%%%%%%%%%%%%%%%%%%%%%%%%%%%%%%%%%%%%%%%%%%%%%%%%%%%%%%%%%%%%%%%%%%
MTF thanks Brazilian Research Agency CNPq for its financial support.  SLB was
supported by the Humboldt Foundation.  MH would like to thank Dr.  Thomas
Richter and Prof.  Harry Paul for useful conversations.

%%%%%%%%%%%%%%%%%%%%%%%%%%%%%%%%%%%%%%%%%%%%%%%%%%%%%%%%%%%%%%%%%%%%%%%%%%%%%%
%                               FIGURE DIRECT
%%%%%%%%%%%%%%%%%%%%%%%%%%%%%%%%%%%%%%%%%%%%%%%%%%%%%%%%%%%%%%%%%%%%%%%%%%%%%%
%

%%%%%%%%%%%%%%%%%%%%%%%%%%%%%%%%%%%%%%%%%%%%%%%%%%%%%%%%%%%%%%%%%%%%%%%%%%%%%%
%                               FIGURE INDIRECT
%%%%%%%%%%%%%%%%%%%%%%%%%%%%%%%%%%%%%%%%%%%%%%%%%%%%%%%%%%%%%%%%%%%%%%%%%%%%%%
%

%%%%%%%%%%%%%%%%%%%%%%%%%%%%%%%%%%%%%%%%%%%%%%%%%%%%%%%%%%%%%%%%%%%%%%%%%%%%%%
%                               FIGURE DISTRIBUTION
%%%%%%%%%%%%%%%%%%%%%%%%%%%%%%%%%%%%%%%%%%%%%%%%%%%%%%%%%%%%%%%%%%%%%%%%%%%%%%
%

%%%%%%%%%%%%%%%%%%%%%%%%%%%%%%%%%%%%%%%%%%%%%%%%%%%%%%%%%%%%%%%%%%%%%%%%%%%%
%                       REFERENCES
%%%%%%%%%%%%%%%%%%%%%%%%%%%%%%%%%%%%%%%%%%%%%%%%%%%%%%%%%%%%%%%%%%%%%%%%%%%%

\end{document}